# Spin transport of electrons and holes in a metal and in a semiconductor

V. Zayets[*]

*Spintronic Research Center, National Institute of Advanced Industrial Science and Technology (AIST), Umezono 1-1-1, Tsukuba, Ibaraki, Japan. e-mail:v.zayets@aist.go.jp*

---

Abstract:

*The features of the spin and charge transport of electrons and holes in a metal and a semiconductor were studied using the Boltzmann transport equations. It was shown that the electrons and holes carry the spin in opposite directions in an electrical current. As result, the spin polarization of an electrical current in a metal is substantially smaller than spin polarization of electron gas. It was shown that the spin properties of the electron gas are responsible for the existence of the concept of "electrons" and "holes" in a metal and a semiconductor.*

---

---

## 1. Introduction

Conventionally, the spin transport in a conductor is described by the model of spin-up/spin-down bands [1]. This model assumes that all electrons of an electron gas can be divided into two independent groups: a group of electrons with the spin-up spin projections and a group of electrons with the spin-down spin projections. It was assumed that there is no exchange of electrons between these groups or the exchange is slow. Therefore, each group of electrons has its own thermo-equilibrium and it is possible to assign different Fermi energies and chemical potentials to each group. Because of the different transport of electrons of spin-up and spin-down spin projections, an electrical current can transfer the spin.

There are several unclear assumptions of this model. Firstly, the reason is unclear why there is no electron exchange between groups of electrons of spin-up spin-down spin projections and why these groups of electrons are thermo isolated. Secondly, it is also unclear why spin-up/spin-down direction is special and why not instead the electrons of the spin-left/spin-right or any other spin projections should be thermo isolated.

Additionally, in the model of spin-up/spin-down bands the spin transport is described by the Helmholtz equitation [1], which simplifies the spin transport only to a simple diffusion of particles. Such an over-simplified description ignores several important facts. The Helmholtz equitation describes the diffusion of the spins without any accompanied diffusion of the charge. This is the case when the diffusion of the spin-polarized electrons in one direction is always exactly equal to the diffusion of spin-unpolarized electrons in the opposite direction. This condition contradicts with some experimental facts [2,3]. In the vicinity of an interface the charge is accumulated along spin diffusion [4]. The measurements of this charge accumulation are often used to estimate the magnitude of a spin current. This effect is called the spin detection [2,3]. It should be noted that the problem of the spin detection can be resolved by modifying the model of spin-up/spin-down bands. In Refs. [5-7] it was assumed that the conductivity of electrons of the spin-up and spin-down bands may be different. In this case the spin detection effect can be described.

Another over-simplified assumption of the model of Ref. [1] is the assumption that all conduction electrons have the same spin-transport properties. This is not correct. The spin-transport properties of electrons with energy higher and lower than the Fermi energy $E_F$ are substantially different. For example, in a n-type semiconductor, where the energy of conduction electrons is higher than $E_F$, the spins are drifted along the movement direction of the electrons. In contrast, in a p-type semiconductor, in which the energy of conduction electrons is lower than $E_F$, the spins are drifted in the opposite direction along the movement of the holes.

The equilibrium distribution of electrons in an electron gas with respect to their energy and spin directions is quickly established and sustained by frequent spin-independent scatterings. The properties of the scatterings significantly influence both the equilibrium spin distribution and the features of the spin transport. The properties of the spin-independent scatterings and their influence on spin properties of the electron gas have been studied in Ref.8. It was shown that the spin direction of each electron may rotate after a scattering. Despite of the frequent spin rotations after scatterings, still electrons of an electron gas can be divided into two groups of spin-polarized and spin-unpolarized electrons, because the spin-independent scatterings do not change the number of electrons in each group [8]. In the group of spin-polarized electrons, all electron spins are directed in one direction. In the group of spin-unpolarized electrons, the spins are equally distributed in all directions. That means that for any chosen direction there is an equal amount of spin-up and spin-down electrons. It is important that the latter condition is valid for any chosen direction. Therefore, the time-inverse symmetry is not broken for the group of spin-unpolarized electrons.

In contrast, in the model of spin-up/spin-down bands the time-inverse symmetry is broken even for the group of spin-unpolarized electrons, which is an incorrect result of this model. This fact is explained as follows. The main assumption of the model of spin-up/spin-down bands is that the transports of electrons of spin-up and spin-down projections are independent. Since the spin-up and spin-down electrons have different transport properties, they can be distinguished by their spin-down/spin-up projection along one fixed axis. The group of spin-unpolarized electrons consists of an equal amount of spin-up and spin-down electrons. Still one fixed axis can be distinguished in this group. That means the time-inverse symmetry is broken even for the group of the spin-unpolarized electrons. This is an incorrect conclusion of the model of spin-up/spin-down bands. The total spin of the group of spin-polarized electrons is zero. Therefore, the time inverse-symmetry of this group should not be broken, there should be no any distinguished direction for this group and all directions should be absolutely equal. For example, in a ferromagnetic metal the time-inverse symmetry of the group of spin-unpolarized electrons is not broken along metal magnetization direction. It is broken only for the group of the spin-polarized electrons. It should be noticed that the model of the spin-up/spin-down bands [1] and the model of Ref. 8 give the same description of the electron gas, when there is no spin or/and charge currents. There are only differences in the description of the spin/charge transport.

In this paper the description of the spin and the charge transport in electron gas includes the following important facts. The description is based on results of the model of Ref. [8], which correctly describe the fact that the total spin of the group of the spin-unpolarized electrons is zero and the time-inverse symmetry is not broken for this group of electrons. Another fact is that the spin-polarized and spin-unpolarized electrons have different energy distributions and this difference influences the transport. The third fact is that influence of different transport mechanisms on the spin transport may be substantially different and the contribution of each transport mechanism to the spin transport should be described separately.

This paper is organized as follows. In Chapter 2 the spin/charge transport equations are obtained from the continuity equations. The obtained spin/charge transport equations contain 5 unknown parameters: 4 different conductivities and the spin life time. In Chapters 3, the Boltzmann transport equations are solved in order to calculate these 4 conductivities for the case of the spin and charge transport in the bulk of a conductor with a low density of defects. In Chapter 4, the properties of the conductivities are analyzed for transport in the bulk of a high-conductivity conductor. The influences of the defects and the interfaces on the conductivities are discussed in Chapter 5. In Chapter 6 the spin-related features of the "electrons" and "holes" in a metal and a semiconductor are described.

## 2. Transport equations

The quantum-mechanical description of the electron transport is based on the following facts. The delocalized (conduction) electrons of the electron gas occupy quantum states, the properties of these quantum states and the electron scatterings between these states determine the electron transport. Each quantum state is distinguished by the direction of its wavevector in the Brillouin zone, its energy and its spatial symmetry. Due to the Pauli excursion principle, each quantum state can be occupied maximum by two electrons of opposite spins. At each moment a quantum state is filled either by no electrons or one electron or two electrons of opposite spins.

As was mentioned in the introduction, all delocalized (conduction) electrons of the electron gas can be divided into two groups of the spin-polarized and spin-unpolarized electrons. The states, which are occupied by two electrons of opposite spins, have zero total spin. All these states belong to the group of spin-unpolarized electrons. A quantum state, which is occupied only by one electron, has spin ½. A part of the half-occupied states belongs to the group of spin-polarized electrons and another part belongs to the group of spin-unpolarized electrons. The important fact is that the time-inverse symmetry is not broken for the group of the spin-unpolarized electrons. It means that the total spin of this group is zero and the projection of the total spin on any axis is zero as well. It is only possible when the spin directions of the states, which are filled only by one electron, are equally distributed in all directions. As was shown in Ref.[8] the frequent spin-independent scatterings make spins equally distributed in all directions for electrons of the group of spin-unpolarized electron and the total spin of this group equal to zero. In the contrast, the spins of all electrons of the group of spin-polarized electrons are directed in one direction. It has been proven in Ref. [8] that the number of electrons in each group is conserved after a spin-independent scattering. It is the reason why it is possible to calculate individually the electron transport for each group of spin-polarized and spin-unpolarized electrons.

The spin polarization *sp* of the electron gas defines the ratio of the number of the electrons in the group of spin-polarized electrons to the total number $n_{spin}$ of states, which are occupied only by one electron:

$$sp = \frac{n_{TIA}}{n_{spin}} = \frac{n_{TIA}}{n_{TIA} + n_{TIS}} \qquad (1)$$

where $n_{TIA}$ is the number spin-polarized electrons and $n_{TIS}$ is the number of states in the group of spin-unpolarized electrons, which are filled by one electron. The full-filled states, which are occupied by two electrons of opposite spins, are not included into the definition of spin polarization (Eq. 1). It is because the conversion between the groups due to different spin relaxation/pumping mechanisms occurs only between states filled by one electron [8]. The half-filled states of both groups of spin-polarized and spin-unpolarized electrons have spin ½, but the spin of a full-filled state is zero. Mainly the spin pumping (spin relaxation) occurs due to alignment (disalignment) of spin directions along one direction (from one direction). The spin alignment or disalignment may occur only for states with a non-zero spin. The spin conservation law limits from the direct conversion of electrons of the full-filled states of zero spin into the group of the spin-polarized electrons.

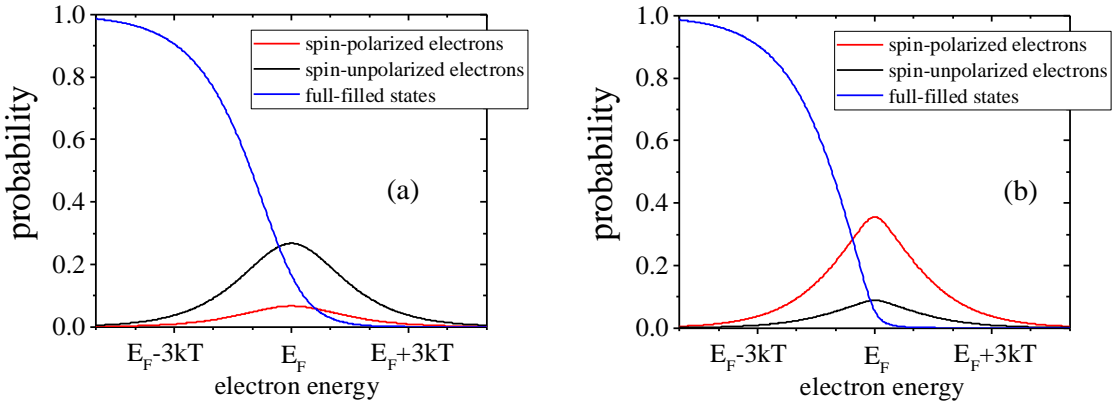

*Figure 1. Probability that a quantum state in electron gas is occupied by an electron from group of spin-polarized electrons (red line) or from group of spin-unpolarized electrons (black line) or the state is occupied by two electrons of opposite spins (blue line). Spin polarization of electron gas is (a) 20 % (b) 80 %*

The numbers of electrons in each group of spin-polarized and spin-unpolarized electrons depend on the electron energy. Nearly all deep-energy states, which energy is substantially below the Fermi energy, are occupied by two electrons and nearly all of them are spin-unpolarized. The states of energy substantially higher the Fermi energy are either not occupied or occupied only by one electron. The energy distributions of electrons in each group can be calculated using the Fermi-Dirac statistics and the features of the spin-independent scatterings [8]. Figure 1 shows the energy distributions of full-filled states, half-filled states, which belong to group of spin-polarized electrons, and half-filled states, which belong to group of spin-polarized electrons. The half-filled states of both groups are mainly distributed in the vicinity of the Fermi energy in the region ± 3 kT. As it is shown in Chapter 3, the shape of the energy distributions of Fig.1 mainly defines the charge and spin transport properties in an electron gas.

The energy distributions are different for spin-polarized and spin-unpolarized electrons because of the different scattering probabilities of the spin-polarized and spin-unpolarized electrons. This fact can be understood as follows. In the case when a state is already occupied by one electron, which spin is up, it is only possible for another electron to be scattered into this state only if its spin direction is spin-down. The spin of the scattered electron should be exactly opposite to the spin of the electron, which is already occupying the state. In the group of the spin-polarized electrons, all electrons have the spins in one direction. Therefore, there are no scatterings between states occupied by spin-polarized electrons. In the group of spin-polarized electrons, the spins are distributed equally in all directions. Therefore, there is a probability between 0 and 1 that a spin-unpolarized electron is scattered into a state occupied either by spin-polarized electron or by another spin-unpolarized electron. A full-filled state, which is occupied by two electrons, has zero spin and no defined spin direction. Therefore, electrons of this state have no spin limitations for a scattering.

The definition of the spin polarization (Eq. 1) includes the total number of the half-filled states $n_{spin}$. The $n_{spin}$ can be calculated by integrating the distributions of Fig.1 with the corresponding density of states of a conductor. For example, in a metal, for which density of states is a constant in the vicinity of the Fermi energy, $n_{spin}$ is can be calculated as [8]:

$$n_{spin} = D \cdot kT \cdot (1.14463 + 0.26055 \cdot sp) \qquad (2)$$

where D is the density of states at the Fermi energy. It should be noticed that $n_{spin}$ only slightly depends on the spin polarization *sp*. It decreases by about 20 % as the spin polarization increases from 0 to 100%. In contrast, $n_{spin}$ may significantly depend on a charge accumulation. For example, in the case of a non-degenerate n-semiconductor the $n_{spin}$ increases exponentially with a linear increase of the Fermi energy. In a metal it may be assumed that $n_{spin}$ only weakly depends on the charge accumulation.

The spin and charge distributions in samples of different geometries can be calculated from the spin and charge transport equations. The spin and charge transport equations are a set of two differential equations with two variables: the chemical potential µ and the spin polarization *sp*. The transport equations can be derived from the continuity equations for the spin and the charge. The continuity equations describe the conservation laws for the spin and the charge. They require that the amount of spin and charge at each point may change only when either electrons are converted between groups of spin-polarized and spin-unpolarized groups or when electrons defuse from a point to a point.

Electrons can be converted from the group of the spin-polarized electrons into the group of the spin-unpolarized electrons because of the spin relaxation. For example, the spin relaxation occurs when each electron has a slightly different precession frequency. Then, during a precession the spins of spin-polarized electrons disalign from one direction. As result, some of them are converted into the group of spin-unpolarized electrons by scatterings [8]. Since the spin relaxation describes a process of disalignment of the spin-polarized electrons from one direction, the rate of the spin relaxation is always proportional to the number of the spin-polarized electrons.

Electrons can be converted back from the group of the spin-unpolarized electrons into the group of the spin-polarized electrons because of the spin pumping. For example, the spin pumping occurs when a magnetic field applied to the electron gas [8]. In this case the spins of electrons align along the magnetic field. As result, some of spin-unpolarized electrons are converted into the group of spin-polarized electrons. Since the spin pumping describes a process of alignment of the spin-unpolarized electrons into one direction, the rate of the spin pumping is proportional to the number of spin-unpolarized electrons of half-filled states. The spin pumping by illumination of a semiconductor by circular-polarized light is an exception. In this case the spin is transformed from a photon to an electron and the electrons from a full-filled state can be directly converted into the group of the spin-polarized electrons. Spin pumping by circular-polarized light is a more complex process and it will not be described in this paper.

Due to the spin relaxation and the spin pumping, the number of the half-field states in the groups of the spin-polarized and the spin unpolarized electrons changes as:

$$\frac{\partial n_{TIA}}{\partial t} = -\frac{\partial n_{TIS}}{\partial t} = -\frac{n_{TIA}}{\tau_{spin}} + \frac{n_{TIS}}{\tau_{pump}} \qquad (3)$$

where $\tau_{spin}$ is the spin relaxation time and $\tau_{pump}$ is the spin pumping time. It should be noted that both $\tau_{spin}$ and $\tau_{pump}$ may slightly depend on the charge accumulation and the spin polarization of the electron gas.

Substituting Eq. (1) into Eq. (3) gives

$$\frac{\partial n_{TIA}}{\partial t} = n_{spin}\left(\frac{1-sp}{\tau_{pump}} - \frac{sp}{\tau_{spin}}\right) \qquad (4)$$

In equilibrium the numbers of the spin-polarized and spin-unpolarized electrons do not change. Therefore, the spin polarization $sp_0$ of the equilibrium can be calculated from the condition:

$$\frac{\partial n_{TIA}}{\partial t} = n_{spin}\left(\frac{1-sp_0}{\tau_{pump}} - \frac{sp_0}{\tau_{spin}}\right) = 0 \qquad (5)$$

which gives

$$\tau_{pump} = \tau_{spin}\frac{1-sp_0}{sp_0} \qquad (6)$$

Substituting Eq. (6) into Eq. (4) gives the conversion rate between spin-polarized and spin-unpolarized electrons as

$$\frac{\partial n_{TIA}}{\partial t} = -\frac{\partial n_{TIS}}{\partial t} = \frac{n_{spin}}{\tau_{spin}}\left(sp_0\frac{1-sp}{1-sp_0} - sp\right) \qquad (7)$$

Next, we calculate the spin and charge currents, which flows along the gradients of the chemical potential $\mu$ and the spin polarization *sp*. As was mentioned above, the model of spin-up/spin-down bands incorrectly assumes that there are two independent chemical potentials for spin-up and spin-down electron bands, there are two independent energy distributions for spin-up and spin-down electrons and there could be two independent currents for spin-up and spin-down electrons [1]. It is an incorrect assumption because the frequent electron scatterings quickly mix up all electrons of all possible spin polarizations ensuring the existence of only one energy distribution for all electrons of different spin distributions. For this reason the spin-polarized and spin-unpolarized electrons always have the same Fermi energy and the same chemical potential $\mu$.

Even though spin-polarized and spin-unpolarized electrons have the same chemical potential, they diffuse independently along the gradient of the chemical potential $\mu$ and the gradient of the spin polarization *sp,* because of their different conductivities. The current of the spin-polarized $\vec{j}_{TIA}$ and current of spin-unpolarized electrons $\vec{j}_{TIS}$ can be expressed as

$$\vec{j}_{TIA} = \frac{1}{q}\left(\sigma_{\mu,TIA}\nabla\mu + \sigma_{sp,TIA} \cdot kT \cdot \nabla sp\right)$$

$$\vec{j}_{TIS} = \frac{1}{q}\left(\sigma_{\mu,TIS}\nabla\mu + \sigma_{sp,TIS} \cdot kT \cdot \nabla sp\right)$$

(8)

where q is the charge of an electron, k is the Boltzmann constant and T is the temperature. The conductivities $\sigma_{\mu,TIA}$ $\sigma_{sp,TIA}$ $\sigma_{\mu,TIS}$ $\sigma_{sp,TIS}$ can be calculated by solving the Boltzmann transport equations (See next chapter). The currents flowing along $\nabla\mu$ are called the drift currents and the currents flowing along $\nabla sp$ are called the diffusion currents [5].

When the spin-polarized electrons defuse, they transport the spin and the charge. In contrast, the spin-unpolarized electrons only transport the charge. Therefore, the spin current $\vec{j}_{spin}$ and charge current $\vec{j}_{charge}$ can be calculated as

$$\vec{j}_{charge} = \vec{j}_{TIA} + \vec{j}_{TIS} = \frac{1}{q}\left(\sigma_{charge}\nabla\mu + \sigma_{detection} \cdot sp \cdot kT \cdot \nabla sp\right)$$

$$\vec{j}_{spin} = \vec{j}_{TIA} = \frac{1}{q}\left(\sigma_{injection} sp \cdot \nabla\mu + \sigma_{spin} \cdot kT \cdot \nabla sp\right)$$

(9)

where the charge, spin, detection and injection conductivities are defined as

$$\sigma_{charge} = \sigma_{\mu,TIS} + \sigma_{\mu,TIA} \qquad \sigma_{detection} = \left(\sigma_{sp,TIS} + \sigma_{sp,TIA}\right)/sp$$

$$\sigma_{injection} = \sigma_{\mu,TIA}/sp \qquad \sigma_{spin} = \sigma_{sp,TIA}$$

(10)

The charge conductivity $\sigma_{charge}$ is the conventional conductivity of the metal, which describes a charge current flow along a gradient of the chemical potential (the Ohm's law). The spin-diffusion conductivity $\sigma_{spin}$ describes the conventional spin diffusion along a gradient of the spin polarization. The injection conductivity $\sigma_{injection}$ is the conductivity of the drift spin current. It defines a spin polarization of an electrical current, which is always smaller or equal to the spin polarization of the electron gas *sp*. When $\sigma_{injection}$ is not zero, electrical current transfers a spin accumulation from one point to another point. In the case when the spin is transformed from one material to another by an electrical current, the effect is called the spin injection [9-16]. The detection conductivity $\sigma_{detection}$ describes charge diffusion along spin diffusion. It causes a charge accumulation along the spin diffusion. Therefore, it makes possible an electrical detection of a spin current [2-4]. It should be noted that the multiplier *sp* is introduced in front of $\sigma_{detection}$ and $\sigma_{injection}$ based on the requirement of the same time-inverse symmetry of each term of Eqs. (9). The conductivities $\sigma_{charge}$, $\sigma_{spin}$ $\sigma_{injection}$ and $\sigma_{detection}$ generally depend on the spin polarization *sp* of the electron gas and a charge accumulation.

The continuity equations for the charge read:

$$\nabla \cdot \vec{j}_{charge} = q \cdot \frac{\partial n}{\partial t} \tag{11}$$

where $\vec{j}_{charge}$ is the charge current and n is the total number of electrons. In a static case the number of electrons does not change $\frac{\partial n}{\partial t} = 0$.

Only electrons of the group of spin-polarized electrons can transfer the spin. Therefore, the continuity equations for the spins read:

$$\nabla \cdot \vec{j}_{spin} = q \cdot \frac{\partial n_{TIA}}{\partial t} \tag{12}$$

where $\vec{j}_{spin}$ is the spin current

Substituting Eqs. (7),(9) into Eqs (11)(12), the *Spin/Charge Transport Equations* are obtained as

$$\nabla \cdot \left( \sigma_{charge} \cdot \nabla \mu + \sigma_{detection} \cdot kT \cdot sp \cdot \nabla sp \right) = 0$$
$$\nabla \cdot \left( \sigma_{injection} \cdot sp \cdot \nabla \mu + \sigma_{spin} \cdot kT \cdot \nabla sp \right) = q^2 \frac{n_{spin}}{\tau_{spin}} \left( sp - \frac{1-sp}{1-sp_0} sp_0 \right) \tag{13}$$

The transport equations (13) are a set of non-linear differential equations, which in general should be solved numerically. However, in some cases it is possible to solve them analytically. In a simple case of a spin diffusion in the bulk of a non-magnetic metal, in which $\nabla \mu = 0$  $\sigma_{detection} = 0$  $sp_0 = 0$ the transport equations (13) are simplified to

$$\nabla \cdot \left( \sigma_{spin} \cdot kT \cdot \nabla sp \right) = q^2 \cdot sp \frac{n_{spin}}{\tau_{spin}} \tag{14}$$

In the case when $\sigma_{spin}$ does not depend on *sp*, Eq. (14) can be simplified to the Helmholtz equation

$$\nabla^2 sp = \frac{sp}{\lambda_{spin}^2} \tag{15}$$

where the spin diffusion length is calculated from Eq. (14) as

$$\lambda_{spin} = \sqrt{\frac{\sigma_{spin} \cdot kT \cdot \tau_{spin}}{n_{spin} \cdot q^2}} \tag{16}$$

The obtained spin and charge transport equations (Eqs. 13) are important to calculate the distributions of the spin/charge accumulations and the spin/ charge currents in different geometries. In a general case, these equations are non-linear and they should be solved numerically. When it is required, the transport equations (Eqs. 13) may be solved together with the Landau-Lifshitz-Gilbert equation [17,18].

## *3. A solution of Boltzmann transport equations for the band current*

The goal of solving the Boltzmann Transport Equations in this manuscript is to calculate the charge $\sigma_{charge}$, spin-diffusion $\sigma_{spin}$, detection $\sigma_{detection}$ and injection $\sigma_{injection}$ conductivities, which are used in the Spin/Charge Transport

Equations (Eqs.(13)). The Boltzmann equations describe a temporal evolution of the distribution function $F(\vec{r},\vec{p},t)$. The distribution function $F(\vec{r},\vec{p},t)$ describes the probability to find an electron at a point $\vec{r}$ with pulse $\vec{p}$ at time t.

We have modified the Boltzmann Transport Equations in order to include several facts, which are essential for the description of the spin and charge transport. The first fact is that the spin-polarized and spin-unpolarized electrons contribute to the transport differently. The reason for this is the different energy distributions of spin-polarized and spin-unpolarized electrons and electrons of the full-filled states (See Fig.1). In order to describe this fact, it necessary to use 3 distribution functions instead of one. We used individual distribution functions $F_{TIA}$ for half-filled states of the group spin-polarized electrons, $F_{TIS}$ for half-filled states of the group spin-unpolarized electrons and $F_{full}$ for the states, which are filled by two electrons of opposite spins. The second fact, which was used in the modified Boltzmann transport equations, is that there is a conversion between groups of spin-polarized and spin-unpolarized electrons, which is induced by different spin relaxation or/and spin pumping mechanisms.

There are several known mechanisms of the electron transport in a solid such as diffusive, ballistic, hopping transport and so on. Here we have divided all transport mechanisms into two groups. The transport related to the electron movement between scatterings is assigned to one group and the transport related to the electron movement due to the scatterings is assigned to another group. The first transport mechanism is defined as the band current. Only conduction band electrons can contribute to this current. The second transport mechanism is defined as the scattering current. Both the localized and delocalized (conduction) electrons contribute to this current. The path of each electron in the phase space can be separated when it is scattered and when it moves between scatterings. Therefore, the contribution of each transport mechanisms should be treated in the Boltzmann transport equation individually and separately.

Including all these facts, the modified Boltzmann transport equation is given as

$$\frac{dF_i}{\partial t} = \left(\frac{dF_i}{\partial t}\right)_{band} + \left(\frac{dF_i}{\partial t}\right)_{Scattering} + \left(\frac{dF_i}{\partial t}\right)_{force} + \left(\frac{dF_i}{\partial t}\right)_{conversion} + \left(\frac{dF_i}{\partial t}\right)_{relaxation} + \left(\frac{dF_i}{\partial t}\right)_{torque} \quad (17)$$

where "i" labels the distribution functions for the group of spin-polarized electrons as "TIA", group of spin-unpolarized electrons as "TIS" and electrons of full-filled states as "full"; $\left(\frac{dF_i}{\partial t}\right)_{band}$ is the term, which describes a change of the distribution function due to the movement of electrons between scatterings; $\left(\frac{dF_i}{\partial t}\right)_{Scattering}$ is the term, which describes the changing of the distribution function due to the movement of electrons due to a scatterings; $\left(\frac{dF_i}{\partial t}\right)_{force}$ is the force term, which describes the change of the distribution function due an external field (for example, an electrical field) ; the conversion term $\left(\frac{dF_i}{\partial t}\right)_{conversion}$ describes the electron conversion between groups of spin-polarized and spin-unpolarized electrons, because of the spin relaxation or the spin pumping; the relaxation term $\left(\frac{dF_i}{\partial t}\right)_{relaxation}$ describes the relaxation of a distribution function to an equilibrium distribution function due to the electron scatterings; $\left(\frac{dF_i}{\partial t}\right)_{torque}$ is the spin-torque current term, which describes a change the distribution function due to flow of a spin-torque current. The spin-torque current flows when the spin direction of spin-polarized electrons is different at different points of a sample [8]. It induces a torque on spins of the conduction electrons, which forces the spins of different regions to align in the same direction.

Except for some special cases, it is safe to assume that an external perturbation (applied electrical and magnetic fields, a thermo gradient, a gradient of spin accumulation and a spin-orbit effective magnetic field) is sufficiently small so that under the perturbation the distribution function only slightly changes from the distribution function in equilibrium. In this case the distribution function can be represented as

$$F_i = F_{i,0} + F_{i,1} \quad (18)$$

where $F_{i,0}$ is the distribution function at an equilibrium and $F_{i,1}$ describes a small deviation from the equilibrium such that for any point of the phase space the following condition is valid:

$$F_{i,0} >> F_{i,1} \tag{19}$$

It could be further assumed that the relaxation of the distribution function into the equilibrium is linearly proportional to a deviation of the distribution function from the equilibrium. Then, the relaxation term can be calculated as

$$\left(\frac{dF_i}{\partial t}\right)_{relaxation} = -\frac{F_i - F_{i,0}}{\tau_k} = -\frac{F_{i,1}}{\tau_k} \tag{20}$$

where $\tau_k$ is the momentum relaxation time. The following explains why $\tau_k$ is the same for both groups of spin-polarized and spin-unpolarized electrons and electrons of the full-filled states. The electrons are constantly scattered between the groups of spin-polarized and spin-unpolarized electrons at a high rate. The electrons of each group are not thermo isolated. Only an amount of electrons in each group is conserved during frequent electron scatterings [8]. Because of mixing of electrons between the groups, the electrons of all groups relax together toward equilibrium and $\tau_k$ should be the same for all groups. The use of the relaxation term in form of Eq. (20) is called the relaxation-time approximation.

In the following, the conversion term $\left(\frac{dF_i}{\partial t}\right)_{conversion}$ is calculated. The spin-polarized electrons are converted into the group of spin-unpolarized electrons, because of the different mechanisms of the spin relaxation. Each spin relaxation mechanism constantly disaligns the spin of each electron of the group of the spin-polarized electrons from the common spin direction of the group. For example, each electron may have a slightly different precession frequency in a magnetic field due to a slightly different the g-factor. Independently on the spin relaxation mechanism, each electrons of the group of the spin-polarized electrons equally contributes to the spin relaxation. For this reason it can be concluded that the rate of the conversion between groups of spin-polarized and spin-unpolarized electrons due to the spin relaxation is proportional to the number of electrons in the group of spin-polarized electrons:

$$\left(\frac{dF_{TIS}}{\partial t}\right)_{conversion} = -\left(\frac{dF_{TIA}}{\partial t}\right)_{conversion} = \frac{F_{TIA}}{\tau_{spin}} \tag{21}$$

where $\tau_{spin}$ is the spin life time.

The spin-unpolarized electrons may be converted into the group of spin-polarized electrons due to the different spin pumping mechanisms. Each spin pumping mechanism constantly aligns the spins of electrons of the group of the spin-unpolarized electrons along one direction. For example, when a magnetic field is applied to the spin-unpolarized electron gas, there is a spin precession around the magnetic field and the spins are aligned along the magnetic field due to the Hilbert damping. The conversion rate due to the spin pumping is linearly proportional to the number of electrons in the group of spin-unpolarized electrons [8]:

$$\left(\frac{dF_{TIA}}{\partial t}\right)_{conversion} = -\left(\frac{dF_{TIS}}{\partial t}\right)_{conversion} = \frac{F_{TIS}}{\tau_{pump}} \tag{22}$$

where $\tau_{pump}$ is the effective spin pump time. As was shown in Ref. [8], $\tau_{pump}$ is inversely proportional to the magnitude of the applied magnetic field.

Combining Eqs. (21) and (22), the conversion term is given as

$$\left(\frac{dF_{TIS}}{\partial t}\right)_{conversion} = -\left(\frac{dF_{TIA}}{\partial t}\right)_{conversion} = \frac{F_{TIA}}{\tau_{spin}} - \frac{F_{TIS}}{\tau_{pump}} \tag{23}$$

The integration of Eq.(23) over all states gives exactly Eq. (4). It means that term $\left(\frac{dF_i}{\partial t}\right)_{conversion}$ describes a local conversion between groups of spin- polarized and spin-unpolarized electrons due to a local spin alignment or disalignment.

It does not describe any changes of the distribution function due to diffusion of electrons from neighbor points in the phase space. All changes of the distribution function due to the diffusion are describes by 3 terms: $\left(\frac{dF_i}{\partial t}\right)_{band}$, $\left(\frac{dF_i}{\partial t}\right)_{Scattering}$, $\left(\frac{dF_i}{\partial t}\right)_{torque}$, which may add additional terms into Eq.(23). For example, a flow of spin-torque current between regions of different spin directions of spin accumulation induces an addition spin relaxation [8].

In this manuscript we only calculate the band current. We assume that the spin direction of the spin-polarized electrons is the same over whole sample and there is no spin-torque-current: $\left(\frac{dF_i}{\partial t}\right)_{torque} = 0$. Also, in this manuscript we neglect the scattering current $\left(\frac{dF_i}{\partial t}\right)_{Scattering} \approx 0$ and we calculate only the band current. The band current is the major transport mechanism in the bulk of a metal and a semiconductor. It is much more efficient than the scattering current. However, in the vicinity of an interface or in a metal with a substantial number of defects the band current decreases and the contribution of the scattering might be essential. Also, the scattering current may flow perpendicularly to the band current. It happens in the case when there are spin-dependent scatterings. The anomalous Hall effect and the Spin Hall effect occur due to the spin-dependent scatterings and a scattering current, which flows perpendicularly to the band current. The scattering current was calculated in Ref.4.

The band current occurs because of the movement of electrons between scatterings. In electron gas the conduction electrons move between scatterings at a high-speed in all directions. In equilibrium, the numbers of electrons moving in any opposite directions are exactly the same. Therefore, there is no electron current. When a voltage is applied to the conductor, the numbers of electrons moving along and opposite to the electrical field become slightly different. Therefore, the band current flows along the electrical field and it transports the charge and the spin.

It is important to emphasize that not all conduction electrons contribute to the band current. All conduction electrons can be divided into two different types: running-wave and standing-wave electrons. The standing-wave electrons do not contribute to the band current. This fact is explained as follows. The effective length of the conduction electrons is rather long (See Fig. 3). In a semiconductor it can be as long as a hundred of nanometers. In the case when an average distance between defects in a conductor is comparable with the effective length of a conduction electron, the electron may bounce back and forwards between defects. It is similar to the case of a photon, which are bouncing between walls of a resonator. When bouncing between defects, one electron, which is moving forward, is firmly fixed to the electron, which is moving backward. These coupled electrons do not move along crystal. They are fixed at position of defects. Importantly, when an electrical field is applied, there is still one electron moving forward and one electron moving backward. The electrical field does not change the ratio of electrons moving in the opposite directions for this type of electrons. Therefore, these electrons do not contribute to the band current. These coupled electrons are defined as the standing-wave electrons. The electrons, which can move freely along crystal and which are not fixed to one position, are defined as the running-wave electrons. Only they contribute to the band current.

There are standing-wave electrons in conductors with defects, in multilayers and in vicinity of the interface between two conductors [4]. The existence of the standing-wave electrons influences significantly the spin and charge transport. For example, in the vicinity of an interface there are more standing-wave electrons and less running-wave electrons comparing to the bulk of the conductor. As result, the conventional conductivity $\sigma_{charge}$ decreases near the interface. The detection $\sigma_{detection}$ and injection $\sigma_{injection}$ conductivities experiences even larger changes. The values of both conductivities are near zero in the bulk, but in the vicinity of interface their values may increase substantially and their values may become comparable to the value of the conventional conductivity $\sigma_{charge}$. It is the reason why the spin detection, which is the effect of charge accumulation along spin diffusion, only has been observed at an interface, but not in the bulk of a conductor [2-4]. It is also the reason why the spin transfer by an electrical current and the spin injection are more effective along or across an interface than in the bulk of a conductor [4].

Below we will solve the Boltzmann transport equations only for the case when there are no any standing-wave electrons. However, in Chapter 4 we will discuss how the existence of the standing-wave electrons influences the obtained results.

As was discussed above, in order to calculate the band current, two terms of the general Boltzmann transport equations have to be used and Eqs. (17) are simplified to:

$$\left(\frac{dF_i}{\partial t}\right)_{band} - \frac{F_{i,1}}{\tau_k} = 0 \tag{24}$$

The solution of Eqs. (24) (See Appendix A) is:

$$\begin{aligned}
\vec{j}_{TIA} &= -\frac{q \cdot \tau_k}{3} \cdot \int D(E) \cdot |\vec{v}|^2 \cdot \nabla F_{TIA,0} \cdot dE \\
\vec{j}_{TIS} &= -\frac{q \cdot \tau_k}{3} \cdot \int D(E) \cdot |\vec{v}|^2 \cdot \nabla F_{TIS,0} \cdot dE \\
\vec{j}_{full} &= -\frac{q \cdot \tau_k}{3} \cdot \int D(E) \cdot |\vec{v}|^2 \cdot \nabla F_{full,0} \cdot dE
\end{aligned} \tag{25}$$

where $\vec{j}_{TIA}, \vec{j}_{TIS}, \vec{j}_{full}$ are the band currents of spin-polarized, spin-unpolarized electrons and electrons of full-filled states, respectively. $F_{TIA,0}, F_{TIS,0}, F_{full,0}$ are equilibrium distribution functions of corresponding groups of electrons.

All three currents transport the charge, but the spin is only transported by the current of the spin-polarized electrons $\vec{j}_{TIA}$. Therefore, the charge current $\vec{j}_{charge}$ and spin current $\vec{j}_{spin}$ can be calculated as

$$\begin{aligned}
\vec{j}_{charge} &= \vec{j}_{TIS} + \vec{j}_{full} + \vec{j}_{TIA} \\
\vec{j}_{spin} &= \vec{j}_{TIA}
\end{aligned} \tag{26}$$

There are only two independent variables: the chemical potential $\mu$ and the gradient the spin polarization *sp*, which describe special variation of the distribution function. Therefore, the gradients of the distribution functions can be calculated as:

$$\nabla F_{i,0} = \frac{\partial F_{i,0}}{\partial \mu} \nabla \mu + \frac{\partial F_{i,0}}{\partial sp} \nabla sp \tag{27}$$

Using Eqs (27) and (25), the charge, injection, spin and detection conductivities are calculated from Eqs. (26) (See Appendix B) as

$$\begin{aligned}
\sigma_{charge} &= \sigma_{\mu,TIA} + \sigma_{\mu,TIS} + \sigma_{\mu,full} \\
\sigma_{injection} &= \frac{1}{sp} \cdot \sigma_{\mu,TIA} \\
\sigma_{detection} &= \frac{1}{sp} \cdot \left(\sigma_{sp,TIA} + \sigma_{sp,TIS} + \sigma_{sp,full}\right) \\
\sigma_{injection} &= \sigma_{sp,TIA}
\end{aligned} \tag{28}$$

The conductivities at the right side of Eqs (28) are calculated as

$$\sigma_j = \frac{q^2 \cdot \tau_k}{3} \cdot \int D(E) \cdot |\vec{v}|^2 \cdot \sigma_j(E) \cdot d(E/kT) \tag{29}$$

where the script j denotes "μ,TIA", " μ,TIS", " μ,full", "sp,TIA", "sp,TIS", "sp,full". The σ$_j$(E) are defined as the state conductivities and are calculated as

$$\sigma_{\mu,TIA}(E) = -kT \frac{\partial F_{TIA,0}}{\partial E} \quad \sigma_{\mu,TIS}(E) = -kT \frac{\partial F_{TIS,0}}{\partial E} \quad \sigma_{\mu,full}(E) = -kT \frac{\partial F_{full,0}}{\partial E}$$
$$\sigma_{sp,TIA}(E) = \frac{\partial F_{TIA,0}}{\partial sp} \quad \sigma_{sp,TIS}(E) = \frac{\partial F_{TIS,0}}{\partial sp} \quad \sigma_{sp,full}(E) = \frac{\partial F_{full,0}}{\partial sp}$$
(30)

The definition of the unitless state conductivities (Eqs.30) does not include the density of state of a conductor. Therefore, common properties of metals or semiconductors can by analyze using the state conductivities.

The obtained conductivities (Eqs. (28)-(30)) should be used in the transport equations (Eqs. 13) in order to calculate the spin/charge transport in different materials. The conductivities were calculated for the case of the band current flowing in the bulk of a conductor without defects. In this simple case the conductivities are proportional to the derivatives of energy distributions of spin-polarized and spin-unpolarized electrons (Fig.1) with respect to the energy and spin polarization (Eqs. 30).

## 4. Properties of charge/spin conductivities $\sigma_{charge}$, $\sigma_{spin}$, $\sigma_{injection}$, and $\sigma_{detection}$ in the bulk of a high-conductivity conductor

Figure 2 shows the calculated charge $\sigma_{charge}$, injection $\sigma_{injection}$, detection $\sigma_{detection}$ and spin-diffusion $\sigma_{spin}$ state conductivities for the spin polarization of electron gas 20% and 85%. The charge or ordinary conductivity $\sigma_{charge}$ (black line) significantly depends on electron energy. The electrons at the Fermi energy are most effective for the transport of the charge. The ordinary charge conductivity for them is largest. Above and below the Fermi energy the ordinary conductivity sharply decreases. For example, the contribution to the ordinary conductivity of electrons at energies 5kT above/below the Fermi energy is near 50 times smaller than the contribution of electrons at the Fermi energy. It is the primary reason why the conductivity of a metal is much higher than the conductivity of a semiconductor. The ordinary conductivity $\sigma_{charge}$ does not depend on the spin polarization *sp* of the electron gas.

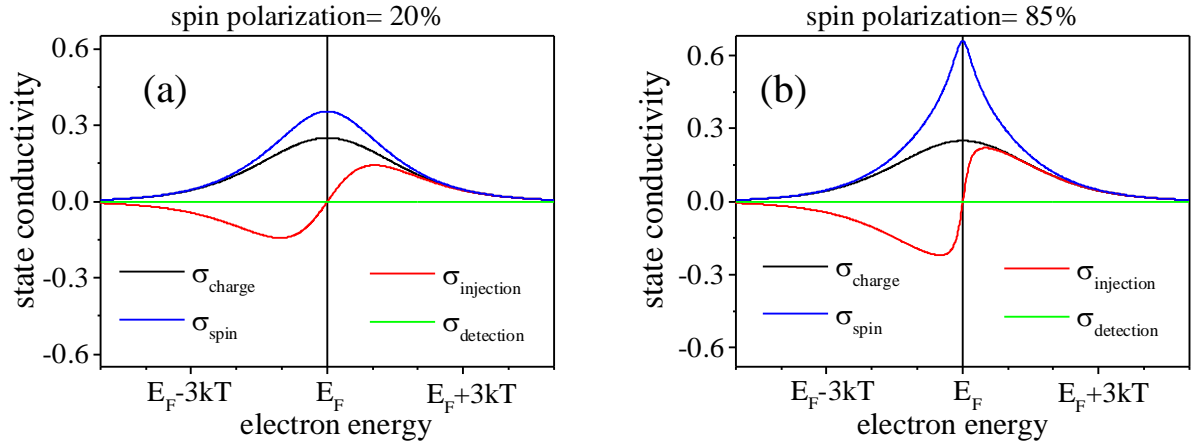

Figure 2. The charge $\sigma_{charge}$, injection $\sigma_{injection}$, detection $\sigma_{detection}$ and spin $\sigma_{spin}$ conductivities for the spin polarization of electron gas (a) *sp*=0.2 ;(b) *sp*=0.85.

The spin-diffusion conductivity $\sigma_{spin}$ (blue line of Fig.2) describes the diffusion of the spins along a gradient of spin accumulation. The spin conductivity $\sigma_{spin}$ is larger than the ordinary charge conductivity $\sigma_{charge}$. Above and below the Fermi energy, these conductivities become near equal. The largest difference between them is at the Fermi energy and it is larger for a larger spin polarization *sp* of the electron gas. At spin polarization sp=75% , the $\sigma_{spin}$ can be 2.5 times larger than $\sigma_{charge}$. The reason of the difference between $\sigma_{spin}$ and $\sigma_{charge}$ is a higher density of spin-polarized electrons at the Fermi energy (See Fig.1).

The injection conductivity $\sigma_{injection}$ (red line of Fig.2) describes the drift of the spins along an electrical current. It also defines the spin polarization of an electrical current. When an electrical current flows from one region to another region, the spin accumulation in one region may increase and in another region it may decrease. When two regions belong to different metals, such effect is called the spin injection. The polarity of the $\sigma_{injection}$ is different for electrons, energy of which is lower and higher than the Fermi energy. This means that electrons of different energies carry the spins in opposite directions. In a metal the electron density is nearly constant at the Fermi energy. Therefore, amounts of electrons, which carry the spin along an electrical current and in the opposite direction, are nearly equal. As result, the spin polarization of electrical current in a metal is near zero and an electrical current nearly does not transport the spin at all. Only in the case when in a metal there is a non-zero gradient of the density of state at the Fermi energy and there is a small difference of amounts of electrons bellow and above the Fermi energy, the value of $\sigma_{injection}$ becomes non-zero and an electrical current may transport the spins. However, in a metal it is still very inefficient. It means that in a metal the spin polarization of electrical current is always substantially smaller than the spin polarization of the electron gas. In contrast, in a semiconductor the $\sigma_{injection}$ is large and it is near equal to the ordinary charge conductivity $\sigma_{charge}$. It also means that in a semiconductor the spin polarization of an electrical current is near equal to the spin polarization of the electron gas. In a non-degenerate n-type semiconductor, the Fermi energy is below the conduction band and all conduction electrons are at energies above the Fermi energy. In this case the spins is transported from a "-" to a "+" electrode. It is similar to the transport of the spin by negatively-charged particles in vacuum, when the spin and the negative charge are carried in the same direction. In a non-degenerate p-type semiconductor, the Fermi energy is above the conduction band and all conduction electrons are at energies below the Fermi energy. In this case the spins are transported from a "+" to a "-" electrode. It is similar to the transport of the spin by positively-charged particles in vacuum, when the spin and the positive charge are carried in the same direction.

The detection conductivity $\sigma_{detection}$ (green line of Fig.2) describes the diffusion of the charge along a gradient of the spin accumulation. When spin-polarized electrons diffuse from a region with a higher spin-accumulation to a region with a lower spin accumulation, additionally there is a current of spin-unpolarized electrons, which flows exactly in the opposite direction. When the opposite currents of the spin-polarized and spin-unpolarized electrons are not equal, the charge is accumulated along the spin-diffusion [4]. This charge accumulation can be measured electrically [2,3]. Since an amount of the charge accumulation is proportional to spin diffusion current, the value the spin diffusion current can be estimated from such electrical measurements [2-4]. As can be seen from Fig.2, the $\sigma_{detection}$ equals to zero for all energies. This means that in the bulk of a defect-free conductor there is always an exact balance between opposite currents of spin-polarized and spin-unpolarized electrons.

## 5. *The conductivities in a conductor with defects and multi-layers*

The above-described properties are only valid for the case of transport in the bulk of a high-conductivity conductor without defects and multi-layers. This is the case when the periodicity of the crystal is not broken and all conduction electrons are running-wave electrons. When the number of defects decreases, the number of the standing-wave electrons increases as well. It significantly changes the properties of the spin transport. When the number of defects increases further, the scattering current become dominated. The spin transport properties of the scattering current are substantially different from that of the band current. In the following we discuss how the existence of the standing-wave electrons influences the conductivities of Fig.2.

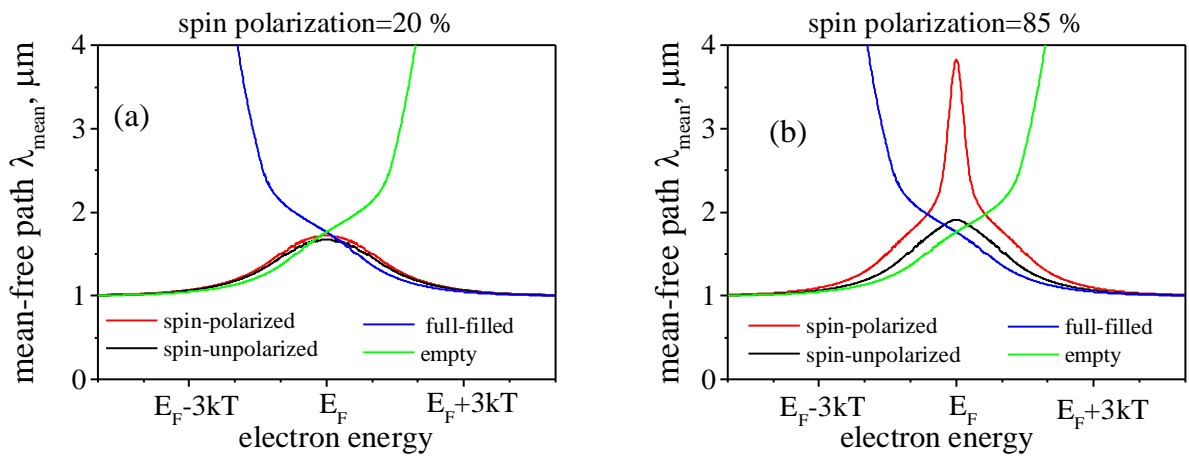

Figure 3. Mean-free path $\lambda_{mean}$ of spin-polarized, spin-unpolarized electrons, electrons of full-filled states and states, which are not filled by any electrons in the bulk of a conductor with defects. The average distance between defects is 1 μm. (a) spin polarization *sp*=0.2; (b) spin polarization *sp*=0.85. There are more standing-wave electrons when $\lambda_{mean}$ is longer.

The mean-free path $\lambda_{mean}$ is an important parameter, which determines the number of the standing-wave electrons in a conductor. The longer the effective length of electron is, the higher the probability is for the electron to bounce between defects or interfaces to form a standing-wave electron. Figure 3 shows the mean-free path $\lambda_{mean}$ of the conduction electrons in the bulk of a conductor with defects, which was calculated in Ref. [4]. The $\lambda_{mean}$ significantly increases for a full-filled states at energies $E_F-2kT$ and lower. Near the Fermi energy, all states have the same and rather short $\lambda_{mean}$. The case of a large spin-polarization *sp*, when the $\lambda_{mean}$ of spin-polarized electrons is longer, is an exception. Intuitively, the data of Fig. 3 can be understood as follows. Below the Fermi energy almost all states are full-filled states, which have no unoccupied places where an electron can be scattered into. There are only a few of half-filled states, which have one unoccupied place. The probability of an electron to be scattered from a full-filled states is very low, because most of time a full-filled state is surrounded by other full-filled states, into which an electron can not be scattered. Rarely a half-filled state is nearby and a scattering event may happen. Therefore, the life time of full-filled state is long and the $\lambda_{mean}$ is long as well. In contrast, a half-state is always surrounded by full-filled states, from which an electron can be scattered into the half-filled state with a high probability. Therefore, both the life time and $\lambda_{mean}$ are short for a half-filled state. Near the Fermi energy, all states have enough possibilities for scatterings and $\lambda_{mean}$ is short for all states, except the case of a high spin polarization *sp* of electron gas. In this case almost all states near the Fermi energy are occupied by spin-polarized electrons. As was mentioned above, electrons can not be scattered between states occupied by spin-polarized electrons. It makes the life time and $\lambda_{mean}$ of the spin-polarized electrons longer.

The existence of the standing-wave electrons affects all four conductivities $\sigma_{charge}$, $\sigma_{spin}$, $\sigma_{injection}$ and $\sigma_{detection}$. Since the standing-wave electrons do not participate in the band current, both the ordinary charge conductivity $\sigma_{charge}$ and the spin-diffusion conductivity $\sigma_{spin}$ decrease. For example, the ordinary conductivity $\sigma_{charge}$ decreases in a conductor with defects and in the vicinity of an interface. Due to the existence of the standing-wave electrons at an interface between two metals, there is always a contact resistance between the metals even in the case when there is no energy barrier between them. When a thickness of a metal decreases, its conductivity $\sigma_{charge}$ becomes smaller because of an increase of the number of the standing-wave electrons. However, when the thickness becomes thinner than the mean-free path $\lambda_{mean}$, the conductivity may become larger again because the transport changes to the 2D type. Because of the electron confinement, the electrons can not be scattered across the interface and they move only in one direction. As result, the number of standing-wave electrons decreases and the conductivities $\sigma_{charge}$ and $\sigma_{spin}$ increase.

As can be seen from Fig.3, the number of the standing-wave electrons significantly dependents on the electron energy. As consequence, the spin properties of the conductivity may change substantially. For example, the detection conductivity $\sigma_{detection}$ may become non-zero. As can be seen from Fig. 3, the mean-free path $\lambda_{mean}$ of full-filled states is rather long for energies below $E_F-2kT$. Therefore, many of electrons of these states are standing-wave electrons and they do not participate in the band current. The reason, why the detection conductivity $\sigma_{detection}$ is zero in Fig.2, is the exact balance of the opposite flows of spin-polarized and spin-polarized electrons along a gradient of spin accumulation. In a conductor with defects or in the vicinity of an interface, more spin-unpolarized electrons and less spin-polarized electrons become the standing-wave electrons. It breaks the balance and the $\sigma_{detection}$ becomes non-zero. This is the reason why the spin detection effect is only observed at a contact between metals [2,3], but not in the bulk of a metal.

The injection conductivity $\sigma_{injection}$ may increase due to the existence of the standing-wave electrons. As was explained above, in an electrical current the electrons of energy above and below the Fermi energy $E_F$ transport the spins in opposite directions. Since a metal has almost the same amounts of electrons with energies above and below the Fermi energy $E_F$, the opposite spin currents nearly compensate each other and the $\sigma_{injection}$ in a metal is small. Such balance may be broken due to standing-wave electrons. The number of standing-wave electrons is not the same for energies above and below the Fermi energy $E_F$ [4]. It makes $\sigma_{injection}$ larger. Such an enlargement of $\sigma_{injection}$ is more profound in the vicinity of an interface [4]. It has an important consequence. Due to the enlargement of the $\sigma_{injection}$ at an interface, the spin injection is more effective through a contact between metals rather than within a single metal. For this reason, the threshold current for magnetization reversal by the spin-transfer torque in a magnetic junction [19,20] is smaller than the threshold current for a current-induced domain movement [21]. In the first case the spin-transfer torque is due to spin injection between metals [6], when the spin injection is more effective. In the second case the spin-transfer torque is due to the spin injection within one metal, when the spin injection is less effective.

In a metal with a high density of defects and a high resistance or in the vicinity of a high-resistance contact, the scattering current may become a dominated transport mechanism. The spin transport properties of the scattering current are substantially different from those of the band current. For example, the ordinary charge conductivity $\sigma_{charge}$ may significantly depend on the spin polarization of the electron gas, which is a rare effect for the drift current. Both the injection conductivity $\sigma_{injection}$ and the detection conductivity $\sigma_{detection}$ are larger for a scattering current. All spin-dependent effects are more profound for a scattering current [4].

## *6. The "electrons" and the "holes"*

It is not only the spin transport, which are influenced by the spin properties of electron gas, but the properties of ordinary conductivity in a non-magnetic conductor significantly depends on the spin features of electron gas. In this section it is shown that the spin properties of the electron gas, which are described in Chapters 2-5, are responsible for the existence of the concept of "electrons" and "holes" in a semiconductor and a metal.

The model of "electrons" and the "holes" in a semiconductor is a very successful model, which describes properties of semiconductors and semiconductor devices. This model assumes that in a semiconductor along negatively-charge particles, the "electrons", there are positively-charged particles, the "holes". All experimental facts well agree with this model. For example, the polarity of the Hall voltage in a p-type semiconductor, where dominated carries are the "holes", is positive. The effect is the same as it should be for a current of the positively-charged particles. However, it is well-known that in a solid the spin and the charge are carried only by negatively-charged electrons. All positive charges are inside of atomic nuclears. The nuclear does not move along the crystal. Therefore, the positive charge of nuclear does not contribute to any charge and spin currents. Therefore, the "hole" is the property of the electron gas, when negatively-charged electrons behave like positively-charged particles.

It could be assumed that the "holes" are voids in the electron gas or unfilled orbitals or some quasi-particles. These assumptions are not correct for the following reasons. The properties of the "holes" are very similar to the properties of the "electrons". Similar to an "electrons", the mean-free path $\lambda_{mean}$ or the effective length of a "hole" may reach hundreds of nanometers in a high-crystal-quality semiconductor. Therefore, the effective length of the "hole" may be equal to the size of thousands of orbitals and a "hole" can not be related to only one orbital. Even though the mobility of the "holes" is slightly smaller than the mobility of the "electrons" in most of semiconductors, there are semiconductors (for example, PbTe, PbS, diamond), in which the hole mobility is larger than the electron mobility. This suggests that the "holes" and the "electrons" are very similar particles.. The "hole" can not be a quasi particle, because the "hole" has the defined spin and magnetic moment.

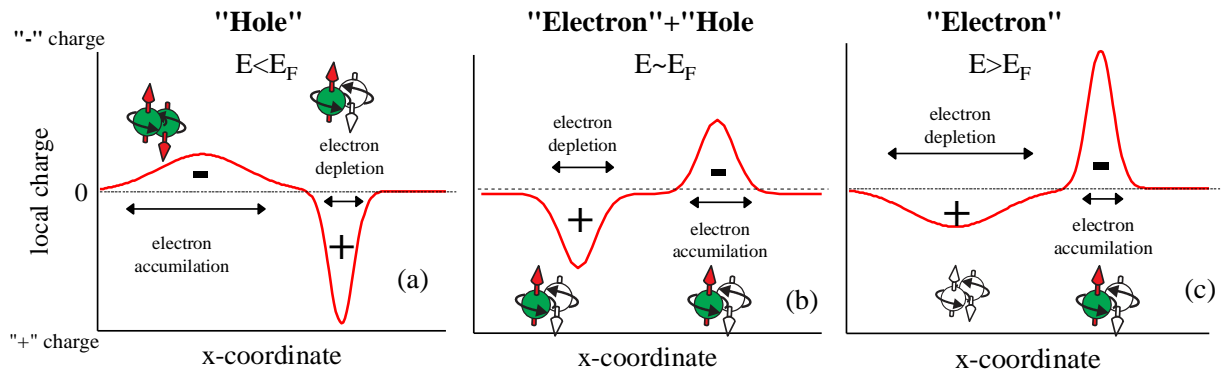

Figure 4. Charge distribution along crystal, when one electron moved aside from an equilibrium position due to a thermo-fluctuation. (a) Positive charge, which is associated with a "hole". Electron energy is lower than the Fermi energy $E<E_F$
(b) There are both "hole" and "electron" when $E\sim E_F$ . (c) Negative charge, which is associated with an "electron". $E>E_F$ .
In all cases position of a spark-like peak of charge accumulation/depletion is at location of a half-filled state.

Also, it is unclear, how to apply the model of "electrons" and the "holes" for a metal. The Hall voltage in a metal is significantly smaller than in a semiconductor. This implies that in a metal there are nearly-equal numbers of the "electrons" and the "holes". The "electrons" and the "holes" have the Hall voltages of opposite polarities, which nearly compensate each other. There are metals, which have a positive or negative Hall voltage. Therefore, a metal can be either "hole"-dominated or "electron"-dominated. It is unclear whether the "electrons" and the "holes" are mixed in a metal or they are separated. It is unclear whether the spins of "electrons" and the "holes" should be in the same direction or their spin directions can be independent. It is unclear whether "electrons" and the "holes" attracts each other to form an exciton as they would a in a semiconductor.

In fact, both the "electrons" and the "holes" are absolutely identical particles. They both are half-filled states, in which one place is occupied by an electron and another place is empty. The only difference between "electrons" and the "holes" is their electron energy. The energy of an "electron" is above the Fermi energy $E_F$ and the energy of a "hole" is below $E_F$. Depending on the electron energy, the properties of a half-filled state are substantially different. As can be seen from Fig.2 and Fig.B1, in an electrical current the "electrons", whose energy $> E_F$ , transport the charge and the spin from a "-" potential to a "+" potential similar to a current of negative-charged particles in vacuum. In contrast, the "holes", whose energy $< E_F$, transfer the charge and the spin in the opposite direction from a "+" potential to a "-" potential similar to a current of positively-charged particles in vacuum. In a metal there are almost equal amounts of "holes" and the "electrons" and they transport spin in the opposite directions. As result, the transferring of the spin accumulation by an electric current is ineffective in the bulk of a metal.

Additionally, a "hole" may be associated with a positive charge and an "electron" may be associated with a negative charge. Figure 4 explains this fact. In equilibrium the charge distribution in a conductor is smooth and it equals zero at any special point. A thermo-fluctuation may move an electron in space (for example, from left to right). Then, at the right side there is a negative charge of electron accumulation and at the left side there is a positive charge of electron depletion.

Figure 4(a) explains why the "hole" is positively charged. At energy lower than $E_F$, a quantum state is filled either by one or by two electrons (See Fig.1). There are no empty states. Therefore, an electron may move only from a full-filled state to a half-filled state. As result, in the region of electron depletion there is an additional half-filled state and in the region of electron accumulation there is an additional full-filled state. The effective length of the half-filled state is significantly shorter than the effective length of the full-filled state (See Fig.3). Therefore, there is a spark-like positively-charged region at the location of the half-filled state, which is defined as the "hole". In contrast, the negatively-charged region at the location of the full-filled state is broad, the magnitude of charge accumulation is small and the region is nearly unnoticeable. Additionally, because of the long effective length, the full-filled state may be bound to a dopant or defect and may not move. In this case only positively-charged half-filled state (the "hole") moves along the crystal.

At energy higher than $E_F$ (Fig.4(c)), a quantum state is either not filled or filled only by one electron (See Fig.1). Therefore, an electron may move only from a half-filled state to an empty state. As result, in the region of electron depletion there is an additional empty state and in the region of electron accumulation there is an additional half-filled state. The effective length of the half-filled state is significantly shorter than the effective length of the empty state (See Fig.3). There is a spark-like negatively-charged region at the location of the half-filled state, which is defined as the "electron". In contrast, the positively-charged region at the location of the empty state is broad and unnoticeable (Fig.4(c)). Therefore, there is a region of a negative charge at the location of the "electron".

At electron energy near $E_F$, all states have a short effective length. Therefore, both the negatively and positively charged regions are sharp (Fig.4(b)). Summing up, at the location of a half-filled state there is a sharp region of positive charge when the electron energy is smaller than $E_F$ and the half-filled state is called the "hole". There is a sharp region of negative charge at the location of a half-filled state when the electron energy is larger than $E_F$ and the half-filled state is called the "electron".

In a metal, the spin-polarized "electrons" and spin-polarized "holes" have the same spin-direction due to frequent electron scatterings between them. In contrast, in a semiconductor the spin-polarized "electrons" and spin-polarized "holes" may have different spin directions. In a semiconductor the "electrons" and "holes" belong to different bands of different symmetries. The "electrons" belong to the conduction band of the s-symmetry and the "holes" belong to the valence band of the p-symmetry. Because of the different symmetries the scatterings between the "electrons" and "holes" are rare. For example, in GaAs an electron is scattered between the conduction and valance bands each 1-30 nanoseconds. In comparison, an electron experiences a scattering between states of the conduction band each 10-100 femtoseconds [22]. Because of rare scatterings between bands, all conduction electrons may be divided into two separated weakly-interacted groups: the electrons of the conduction band (the "electrons") and the electrons of the valence band (the "holes"). The electrons of each group may have their own thermo equilibrium, their own spin direction, their own chemical potential and their own Fermi energy. Because of this unique property of a semiconductor it is possible to fabricate a semiconductor laser and a bipolar transistor.

Similar as electrons are divided into the "holes" and the "electrons, the model of the spin-up/spin-down bands incorrectly assumes that it is possible to use the same method in order to divide all electrons into the groups of spin-polarized and spin-unpolarized electrons or electrons of spin-up and spin-down spin projections. The division into the groups of "electrons" and "holes" in a semiconductor is justified because of a rare exchange of electrons between these groups. The reason, why it is possible to divide electrons into the groups of spin-polarized and spin-unpolarized electrons, is different. On the contrary, the division is possible not because of the rare electron exchange, but because of the frequent electron exchange between the groups of spin-polarized and spin-unpolarized electrons. Because of the frequent electron scatterings and the conservation of the time-inverse symmetry, the number of electrons in each group and the distributions of spin directions in each group do not change for a relatively-long time [8]. There is only one thermo equilibrium for both groups. The introduction of different chemical potentials and the Fermi energies for the spin-polarized and spin-unpolarized electrons or electrons of spin-up and spin-down spin projections has no physical meaning.

## *6. Conclusion*

The spin and charge transport equations (Eqs. 13) were derived. By solving these equations (for example, numerically), the distributions of the spin/ charge accumulations and the spin/ charge currents can be calculated in different geometries. The transport equations include 4 different conductivities, which were calculated by solving the Boltzmann transport equations. The conductivities were calculated and studied for the case of the band current flowing in the bulk of a conductor. In this case the conductivities are simply proportional to the derivatives of energy distributions of spin-polarized and spin-unpolarized electrons (Fig.1) with respect to the energy and spin polarization (Eqs. 30).

It was shown that the "electrons" and "holes" transport the spin in opposite directions in an electrical drift current. Since in a metal there are nearly equal amounts of "electrons" and "holes", the spin polarization of an electrical current in a metal is substantially smaller than the spin polarization of an electron gas and the spin transfer is not effective in the bulk of metal.

It was shown that in the bulk of a conductor the diffusion of spin-polarized electrons is exactly compensated by opposite diffusion of spin-unpolarized electrons. As result, the charge is not accumulated along the spin current and the spin detection effect does not exist in the bulk of a conductor.

It was shown that the spin properties of the electron gas are responsible for the existence of the concept of "electrons" and "holes" in a conductor. It was shown that the "electrons" and the "holes" in a semiconductor and a metal represent a same particle: a half-filled quantum state. The half-filled state is a quantum state of conduction electrons, which is occupied only by one electron. The only difference between an "electron" and a "hole" is their energy with respect to the Fermi energy. The "hole" behaves similar to a positively-charged particle in vacuum and the "electron" behaves similar to a negatively-charged particle in vacuum. The direction, in which the spin and the charge are carried in an electrical current, is along of movement of positively-charged particles in vacuum in the case of a "hole" current. It is along of movement of negatively-charged particles in vacuum in the case of a "electron" current. Additionally, it was shown that there is a narrow region of positive charge at location of a "hole" and there is a narrow region of negative charge at location of an "electron".

# **Appendix A**

Below the Boltzmann transport equations (Eqs.24) are solved for the band current.

The band current occurs, because of the movement of band electrons (delocalized conduction electrons) in space. The movement of an electron literally means that if at time t the electron is at point x, at time t+dt the electron will be at point $x + v_x \cdot dt$, where $v_x$ is the x-axis projection of the electron speed. The change of the electron distribution function F(x) due to the movement of the conduction electrons along the x-direction can be described as:

$$dF(x) = F(x - v_x \cdot dt) - F(x) \tag{A.1}$$

In the case of a short time interval dt, Eq. (A.1) can be simplified as

$$\frac{\partial F}{\partial t} = -v_x \cdot \frac{\partial F}{\partial x} \tag{A.2}$$

Taking into the account that an electron can move not only in the x-direction, but in any direction gives the band-current term of the Boltzmann equation as

$$\left(\frac{\partial F}{\partial t}\right)_{band} = -\vec{v} \cdot \nabla F \tag{A.3}$$

Substituting Eq. (A.3) into the Boltzmann transport equations Eqs. (24) gives

$$-\frac{F_{i,1}}{\tau_k} - \vec{v} \cdot \nabla\left(F_{i,0} + F_{i,1}\right) = 0 \tag{A.4}$$

The solution of Eq. (A.4) using the condition (Eq. 19) is

$$F_{i,1} = -\tau_k \cdot \vec{v} \cdot \nabla F_{i,0} \tag{A.5}$$

The current due to movement of one electron in volume V is

$$\vec{j}_{one\_el} = -\frac{q \cdot \vec{v}}{V} \tag{A.6}$$

where q is the charge of an electron

Integrating Eq. (A.6) over all states and using Eq. (A.5), the band current can be calculated as

$$\vec{j}_i = \frac{q}{(2\pi\hbar)^3}\iiint \vec{v} \cdot (F_{i,0} + F_{i,1}) \cdot d\vec{p} = \frac{q}{(2\pi\hbar)^3}\iiint \vec{v} \cdot F_{i,1} \cdot d\vec{p} = \frac{-\tau_k q}{(2\pi\hbar)^3}\cdot \iiint \vec{v} \cdot (\vec{v} \cdot \nabla F_{i,0}) \cdot d\vec{p} \tag{A.7}$$

In equilibrium there is no current, this can be described as:

$$\frac{q}{(2\pi\hbar)^3}\iiint \vec{v} \cdot F_{i,0} \cdot d\vec{p} = 0 \tag{A.8}$$

The condition (A.8) was used to simplify Eqs. (A.7).

In the case of a transport in the bulk of an isotropic metal, Eq. (A.7) can be further simplified. We define the angle θ as the angle between the electron movement direction and $\nabla F_{i,0}$. Then, the number of electrons in each group of spin-polarized and spin-unpolarized electrons can be calculated as

$$n_i = \frac{1}{(2\pi\hbar)^3}\iiint F_{i,0}(E)\cdot d\vec{p} = \int dE \cdot D(E) \cdot F_{i,0}(E) \cdot \int_0^\pi 0.5\cdot \sin(\theta)\cdot d\theta = \int D(E)\cdot F_{i,0}(E)\cdot dE \tag{A.9}$$

where D(E) is the density of the states.

Similarly, the currents flowing along and perpendicularly to $\nabla F_{i,0}$ are calculated from Eq. (A.7) as

$$j_{i,\parallel} = -\tau_k q \cdot \int D(E)\cdot dE \cdot \int_0^\pi 0.5\cdot \sin(\theta)\cdot d\theta \cdot |\vec{v}|\cdot \cos(\theta)\cdot \left(|\vec{v}|\cdot |\nabla F_{i,0}|\cdot \cos(\theta)\right) = -\frac{q\cdot \tau_k}{3}\cdot \int D(E)\cdot |\vec{v}|^2 \cdot |\nabla F_{i,0}| dE$$

$$j_{i,\perp} = -\tau_k q \cdot \int D(E)\cdot dE \cdot \int_0^\pi 0.5\cdot \sin(\theta)\cdot d\theta \cdot |\vec{v}|\cdot \sin(\theta)\cdot \left(|\vec{v}|\cdot |\nabla F_{i,0}|\cdot \cos(\theta)\right) = 0$$

$$\tag{A.10}$$

Therefore, the current flows only along $\nabla F_{i,0}$ and from Eq. (A.10) the total band current can be calculated as

$$\vec{j}_i = -\frac{q\cdot \tau_k}{3}\cdot \int D(E)\cdot |\vec{v}|^2 \cdot \nabla F_{i,0}\cdot dE \tag{A.11}$$

The explicit expressions for the band current of spin-polarized electrons $\vec{j}_{TIA}$, the band current of spin-unpolarized electrons $\vec{j}_{TIS}$ and the band current of electrons occupying the full-filled states $\vec{j}_{full}$ are

$$\vec{j}_{TIA} = -\frac{q \cdot \tau_k}{3} \cdot \int D(E) \cdot |\vec{v}|^2 \cdot \nabla F_{TIA,0} \cdot dE$$

$$\vec{j}_{TIS} = -\frac{q \cdot \tau_k}{3} \cdot \int D(E) \cdot |\vec{v}|^2 \cdot \nabla F_{TIS,0} \cdot dE \quad (A.12)$$

$$\vec{j}_{full} = -\frac{q \cdot \tau_k}{3} \cdot \int D(E) \cdot |\vec{v}|^2 \cdot \nabla F_{full,0} \cdot dE$$

where $F_{TIA,0}, F_{TIS,0}, F_{full,0}$ are equilibrium energy distributions of spin-polarized, spin-unpolarized electrons and electrons filling full-filled states, which are shown in Fig.1.

It should be noted that simplification from Eq. (A.9) to Eq. (A.11) is only possible, when all conduction electrons are running-wave electrons. In the case of a transport, when there is a substantial amount of standing-wave electrons (for example, in the vicinity of an interface), the integration of Eq. (A.9) is more complex [4].

## *Appendix B*

In the following we calculate the charge $\sigma_{charge}$, spin-diffusion $\sigma_{spin}$, detection $\sigma_{detection}$ and injection $\sigma_{injection}$ conductivities from Eqs. (25) using Eqs. (24), (27) and (9).

In order to simplify the solution and the analysis, two cases are calculated separately. At first, the drift current flowing along an electrical field is calculated. Next, the diffusion current flowing along a gradient of spin polarization is calculated.

In the case when there is a spatial gradient of the chemical potential μ, the gradient of the distribution function can be calculated as

$$\nabla F_{i,0} = \frac{\partial F_{i,0}}{\partial \mu} \nabla \mu \quad (B.1)$$

Substituting Eqs. (B.1) into Eqs. (25) gives

$$\vec{j}_{\mu,i} = \nabla \mu \cdot \frac{q \cdot \tau_k}{3} \cdot \int D(E) \cdot |\vec{v}|^2 \cdot \frac{\partial F_{i,0}}{\partial \left(\frac{E}{kT}\right)} \cdot d\left(\frac{E}{kT}\right) \quad (B.2)$$

Comparison of Eq. (B.2) with a definition of the drift current of Eq.(8) gives the conductivities of spin polarized electrons $\sigma_{\mu,TIA}$, spin-unpolarized electrons $\sigma_{\mu,TIS}$ and electrons of full-filled states $\sigma_{\mu,full}$ for the *drift* current as

$$\sigma_{\mu,TIA} = \frac{q^2 \cdot \tau_k}{3} \cdot \int D(E) \cdot |\vec{v}|^2 \cdot \sigma_{\mu,TIA}(E) \cdot d\left(\frac{E}{kT}\right)$$

$$\sigma_{\mu,TIS} = \frac{q^2 \cdot \tau_k}{3} \cdot \int D(E) \cdot |\vec{v}|^2 \cdot \sigma_{\mu,TIS}(E) \cdot d\left(\frac{E}{kT}\right) \quad (B.3)$$

$$\sigma_{\mu,full} = \frac{q^2 \cdot \tau_k}{3} \cdot \int D(E) \cdot |\vec{v}|^2 \cdot \sigma_{\mu,full}(E) \cdot d\left(\frac{E}{kT}\right)$$

where the state conductivities $\sigma_{\mu,TIA}(E)$, $\sigma_{\mu,TIS}(E)$, $\sigma_{\mu,full}(E)$ are defined as

$$\sigma_{\mu,TIA}(E) = -kT \frac{\partial F_{TIA,0}}{\partial E} \quad \sigma_{\mu,TIS}(E) = -kT \frac{\partial F_{TIS,0}}{\partial E} \quad \sigma_{\mu,full}(E) = -kT \frac{\partial F_{full,0}}{\partial E} \quad (B.4)$$

All conductivities of Eqs. (B.4), which describe a drift current, are proportional to derivative of the energy distribution shown in Fig.1 with respect to the electron energy.

Figure B.1(a) shows the calculated state conductivities σ_{μ,TIA}(E), σ_{μ,TIS}(E), σ_{μ,full}(E). The state conductivity σ_{μ,full}(E) for a band current of electrons of full-filled states is positive for all energies. This means that the full-filled states are drifted from a "–" source toward a "+" drain. It is similar to the movement of a negatively-charged particle in an electrical field in vacuum. The state conductivities σ_{μ,TIA}(E), σ_{μ,TIS}(E) for a current of electrons of half-filled states are positive for energies above the Fermi energy and they are negative for energies below the Fermi energy. This means that the drift direction of these states depends on their energy. The half-filled states of energies above the Fermi energy are drifted from a "–" source toward a "+" drain. It is similar to the movement of a negatively-charged particle in an electrical field in vacuum. However, the half-filled states of energies below the Fermi energy are drifted in the opposite direction from a "+" source toward a "-" drain. The drift becomes similar to the movement of a positively-charged particle in an electrical field in vacuum.

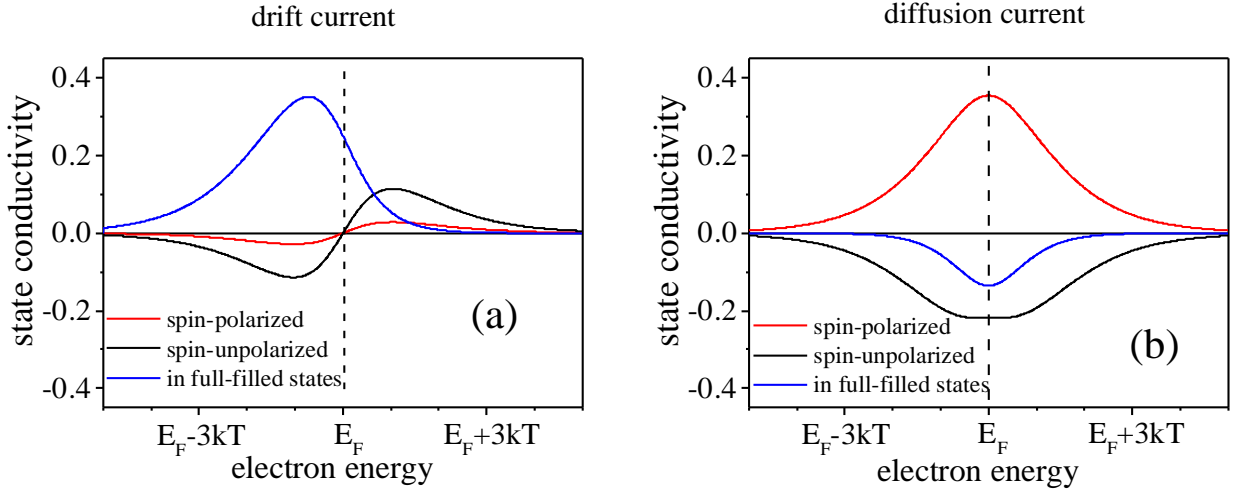

*Fig. B.1. (a) State conductivities $\sigma_{\mu,TIA}(E)$ (red line), $\sigma_{\mu,TIS}(E)$ (black line), $\sigma_{\mu,full}(E)$ (blue line) for a drift current, which flows along an applied electrical field. (b) State conductivities $\sigma_{sp,TIA}(E)$ (red line), $\sigma_{sp,TIS}(E)$ (black line), $\sigma_{sp,full}(E)$ (blue line) for a diffusion current, which flows along a gradient of spin accumulation. Spin polarization of the electron gas is 20 % in both cases. The calculated state conductivities are derivatives of the energy distributions of Fig.1 with respect of energy and spin polarization.*

Next, the second case is calculated when a diffusion current flows along a gradient of spin accumulation. In this case there is a spatial gradient of the spin polarization **sp** and the gradient of the distribution function can be calculated as

$$\nabla F_{i,0} = \frac{\partial F_{i,0}}{\partial sp} \nabla sp \qquad (B.5)$$

Substituting Eqs. (B.5) into Eqs. (25) gives

$$\vec{j}_{sp,i} = \nabla sp \cdot kT \cdot \frac{q \cdot \tau_k}{3} \cdot \int D(E) \cdot |\vec{v}|^2 \cdot \frac{\partial F_{i,0}}{\partial sp} \cdot d\left(\frac{E}{kT}\right) \qquad (B.6)$$

Comparison of Eq. (B.6) with a definition of the drift current of Eq.(8) gives the conductivities of spin polarized electrons $\sigma_{sp,TIA}$, spin-unpolarized electrons $\sigma_{sp,TIS}$ and electrons of full-filled states $\sigma_{sp,full}$ for the *diffusion* current as

$$\sigma_{sp,TIA} = \frac{q^2 \cdot \tau_k}{3} \cdot \int D(E) \cdot |\vec{v}|^2 \cdot \sigma_{sp,TIA}(E) \cdot d\left(\frac{E}{kT}\right)$$

$$\sigma_{sp,TIS} = \frac{q^2 \cdot \tau_k}{3} \cdot \int D(E) \cdot |\vec{v}|^2 \cdot \sigma_{sp,TIS}(E) \cdot d\left(\frac{E}{kT}\right) \qquad (B.7)$$

$$\sigma_{sp,full} = \frac{q^2 \cdot \tau_k}{3} \cdot \int D(E) \cdot |\vec{v}|^2 \cdot \sigma_{sp,full}(E) \cdot d\left(\frac{E}{kT}\right)$$

where the state conductivities $\sigma_{sp,TIA}(E)$, $\sigma_{sp,TIS}(E)$, $\sigma_{sp,full}(E)$ are defined as

$$\sigma_{sp,TIA}(E) = \frac{\partial F_{TIA,0}}{\partial sp} \quad \sigma_{sp,TIS}(E) = \frac{\partial F_{TIS,0}}{\partial sp} \quad \sigma_{sp,full}(E) = \frac{\partial F_{full,0}}{\partial sp} \qquad (B.8)$$

All conductivities of Eq.B.8, which describe a diffusion current, are proportional to derivative of the energy distribution shown in Fig.1 in respect to the spin polarization *sp*.

Figure B.1(b) shows the calculated state conductivities $\sigma_{sp,TIA}(E)$, $\sigma_{sp,TIS}(E)$, $\sigma_{sp,full}(E)$. The conductivity of the spin-polarized electrons $\sigma_{sp,TIA}(E)$ is positive. It describes the simple fact that the spin-polarized electrons diffuse from a region of a higher spin accumulation into a region of a smaller spin accumulations. The conductivities of spin-unpolarized electrons $\sigma_{sp,TIS}(E)$, $\sigma_{sp,full}(E)$ are negative. It means that the spin-unpolarized electrons diffuse in the opposite direction.

The charge is transported by all currents of spin-polarized, spin-unpolarized electrons and electrons of full-filled states, but the spin is only transported by the spin-polarized electrons. Therefore, the charge current $\vec{j}_{charge}$ and spin current $\vec{j}_{spin}$ can be calculated as

$$\vec{j}_{charge} = \vec{j}_{\mu,TIS} + \vec{j}_{\mu,full} + \vec{j}_{\mu,TIA} + \vec{j}_{sp,TIS} + \vec{j}_{sp,full} + \vec{j}_{sp,TIA}$$
$$\vec{j}_{spin} = \vec{j}_{\mu,TIA} + \vec{j}_{sp,TIA}$$
(B.9)

Substituting Eqs.(B.3), (B.7) into Eqs (B.9) and comparing it with Eqs. (8), the charge, injection, spin-diffusion and detection conductivities are calculated as

$$\sigma_{charge} = \sigma_{\mu,TIA} + \sigma_{\mu,TIS} + \sigma_{\mu,full}$$
$$\sigma_{injection} = \frac{1}{sp} \cdot \sigma_{\mu,TIA}$$
$$\sigma_{detection} = \frac{1}{sp} \cdot \left( \sigma_{sp,TIA} + \sigma_{sp,TIS} + \sigma_{sp,full} \right)$$
$$\sigma_{injection} = \sigma_{sp,TIA}$$
(B.10)